\documentclass[
  aps,
  prd,
  reprint,
  superscriptaddress,
  nofootinbib,
  longbibliography,
  floatfix
]{revtex4-2}

\usepackage[UTF8,scheme=plain]{ctex}
\usepackage{amsmath,amssymb,bm,mathtools,mathrsfs}
\usepackage{mathrsfs}  
\usepackage{graphicx}

\usepackage{subcaption}
\usepackage{tikz}
\usetikzlibrary{calc}
\usepackage{booktabs,array}
\usepackage{xcolor}
\usepackage[hidelinks]{hyperref}
\usepackage{microtype}

\allowdisplaybreaks[4]
\newcommand{\dd}{\mathrm d}
\newcommand{\ii}{\mathrm i}
\newcommand{\ee}{\mathrm e}
\newcommand{\mpl}{M_{\rm Pl}}
\newcommand{\Hc}{\mathcal H}
\newcommand{\kUV}{k_{\rm UV}}

\newcommand{\eRD}{\mathrm{eRD}}
\newcommand{\eMD}{\mathrm{eMD}}
\newcommand{\lRD}{\mathrm{lRD}}
\newcommand{\delD}{\delta_{\rm D}}

\newcommand{\vect}[1]{\bm{#1}}
\newcommand{\avg}[1]{\left\langle #1\right\rangle}
\newcommand{\conavg}[1]{\left\langle #1\right\rangle_{c}}

\newcommand{\intk}{\int\!\frac{\dd^3k}{(2\pi)^3}}

\newcommand{\safegraphics}[2][\linewidth]{%
  \IfFileExists{#2}{\includegraphics[width=#1]{#2}}{%
    \fbox{\parbox[c][4.0cm][c]{0.92\linewidth}{\centering\small
      Figure placeholder\\[0.45em]{\footnotesize\nolinkurl{#2}}\\[0.45em]
      The final figure will be inserted automatically once it is placed in the TeX directory.}}}}

\begin{document}

\title{Effects of Poisson Isocurvature Perturbations on Induced Gravitational Waves}

\author{Zi-An Zhou}
\email{zhouzian@itp.ac.cn}
\affiliation{Institute of Theoretical Physics, Chinese Academy of Sciences (CAS), Beijing 100190, China}
\affiliation{School of Physical Sciences, University of Chinese Academy of Sciences, Beijing 100049, China}

\author{Cheng-Jun Fang}
\email{fangchengjun@itp.ac.cn}
\affiliation{Institute of Theoretical Physics, Chinese Academy of Sciences (CAS), Beijing 100190, China}
\affiliation{School of Physical Sciences, University of Chinese Academy of Sciences, Beijing 100049, China}

\author{Zong-Kuan Guo}
\email{guozk@itp.ac.cn}
\affiliation{Institute of Theoretical Physics, Chinese Academy of Sciences (CAS), Beijing 100190, China}
\affiliation{School of Physical Sciences, University of Chinese Academy of Sciences, Beijing 100049, China}
\affiliation{School of Fundamental Physics and Mathematical Sciences, Hangzhou Institute for Advanced Study, University of Chinese Academy of Sciences, Hangzhou 310024, China}

\begin{abstract}
A sudden transition from an early matter-dominated era to a radiation-dominated era provides a representative mechanism for enhancing induced gravitational waves (GWs). In the primordial black hole (PBH) evaporation scenario, discrete PBH number-density fluctuations naturally generate Poisson isocurvature perturbations. By evaluating the GWs induced by these perturbations through the sudden evaporation transition, we show how the full hierarchy of intrinsic non-Gaussian statistics of a Poisson isocurvature source enters the induced GW spectra. We find that the tensor power spectrum retains only the two-point correlations of the scalar source, while higher-order correlations first appear in the tensor bispectrum. We also find a nontrivial scale dependence in the relative amplitudes of different bispectrum configurations, which may provide additional information for distinguishing the discrete PBH scenario from other induced-GW mechanisms.
\end{abstract}


\maketitle

\section{Introduction}
\label{sec:intro}

Observations of gravitational waves (GWs) have opened a new window on
the physics of the early Universe. Among the possible cosmological
sources, scalar-induced GWs, generated at second order by scalar
perturbations, have attracted considerable attention
\cite{
	Ananda2007,
	Baumann2007,
	BartoloMatarreseRiottoVaihkonen2007,
	KohriTerada2018,
	DomenechReview2021}.
Broadly speaking, there are two main routes to enhancing the induced GW signal. One is to enhance the primordial scalar perturbations, as occurs in many scenarios of primordial black hole (PBH) formation\cite{SasakiEtAl2018,CaiPiSasaki2019}. The other is to modify the subsequent cosmic evolution through a nonstandard expansion history, thereby enhancing the tensor response~\cite{AssadullahiWands2009}. A representative example of the
latter is the sudden-transition mechanism, in which a rapid transition
from an early matter-dominated (eMD) era to a late radiation-dominated
(lRD) era amplifies the induced tensor modes~\cite{
	InomataSudden2019,
	InomataGradual2019,
	PearceEtAl2024,
	ZengFangGuo2025,
	InomataKohriTerada2026}.

PBH evaporation provides a concrete realization of this mechanism:
after PBHs come to dominate the energy density of the Universe, Hawking
evaporation ends the eMD era, reheats the Universe, and drives a rapid
return to radiation domination
\cite{
	Inomata2020,
	PapanikolaouVenninLanglois2021,
	DomenechLinSasaki2021,
	DomenechTrankle2025}. The number-density fluctuations of discrete PBHs provide a Poisson isocurvature source with a white-noise power spectrum. At the two-point level, however, this behavior is not unique to a discrete Poisson source; such a \(k^3\) infrared scaling of the dimensionless spectrum also arises generically for causal sources with a finite correlation length ~\cite{HookMarquesTavaresRacco2021}. The two-point spectrum alone therefore does not uniquely identify the underlying discrete process, motivating us to examine higher-order statistics.

Going beyond the two-point level naturally brings higher-order correlations into consideration. For a non-Gaussian stochastic field, connected higher-point correlation functions contain information that is not captured by the power spectrum and can encode additional information about the underlying physical processes. In cosmology, such non-Gaussian statistics provide an important probe of the physics of the early Universe, including high-energy particle content and interactions~\cite{ChenWang2010,ArkaniHamedMaldacena2016, LeeBaumannPimentel2016}. 
    
For scalar-induced GWs, this information can in principle be imprinted on tensor correlation functions, since the induced tensor perturbation depends quadratically on the scalar source. Motivated by this possibility, previous studies have extensively investigated non-Gaussianity in the scalar source, particularly for primordial curvature perturbations. These include local-type expansions controlled by \(f_{\rm NL}\), \(g_{\rm NL}\), or \(\tau_{\rm NL}\), as well as specific nonlinear mappings and non-Gaussian probability distributions
\cite{
	AdsheadLozanovWeiner2021,
	PernaEtAl2024,
	LiEtAl2024,
	LiAngular2024,
	LiArbitraryOrder2025}.
Related studies have also extended this framework to more general source statistics beyond finite-order local expansions~\cite{IovinoEtAl2026,ZengNingCaiWang2025}.

The Poisson case, however, realizes source non-Gaussianity in a
qualitatively different way. For a homogeneous Poisson point process,
the full hierarchy of connected correlations follows directly from the
discreteness of the underlying population, with their relative
amplitudes fixed by the mean number density. Previous studies of PBH-isocurvature-induced GWs have primarily incorporated the statistical properties of the PBH population through the source two-point function or an effective power spectrum. For example, effects such as an extended PBH mass function or PBH clustering modify the density-perturbation power spectrum and thereby alter the induced GW power spectrum ~\cite{
	PapanikolaouVenninLanglois2021,
	DomenechLinSasaki2021,
	Papanikolaou2022,
	PapanikolaouEtAl2024NG,
	HeEtAl2024,
	DomenechTrankle2025,
	DomenechTrankleIsocurvature2025,
	ZengIsocurvatureLattice2026}.

By contrast, here we retain the full intrinsic Poisson hierarchy of the source, and ask how it enters the induced tensor statistics as independent contributions. At the level of the tensor power spectrum, we find that the higher-order information is invisible after the tensor projection, so the result depends only on the source two-point function. The situation changes at the tensor-bispectrum level. Even for a Gaussian scalar source, the quadratic scalar-to-tensor conversion generates a nonzero tensor bispectrum through Wick contractions, reflecting the nonlinearity of the GW generation process rather than non-Gaussianity of the source. For a Poisson source, the connected source three-point correlations generate an additional contribution to the tensor bispectrum, which carries information about the intrinsic non-Gaussianity of the source and therefore lifts the two-point degeneracy in principle. 
Numerically, however, the contribution that directly carries the
intrinsic Poisson non-Gaussianity remains subleading, limiting the practical efficiency of the bispectrum as a discriminator between a discrete Poisson source and a Gaussianized source with the same two-point spectrum. We also find a nontrivial scale dependence in the relative amplitudes of the benchmark equilateral and squeezed configurations.

The paper is organized as follows. In Sec.~II, we introduce the PBH
reheating background, the Poisson isocurvature source, and the scalar
transfer functions relevant to the evaporation transition. In Sec.~III,
we derive the tensor power spectrum and bispectrum from the Poisson
expansion and present the transverse-traceless (TT) selection rule that determines which
contraction topologies survive. In Sec.~IV, we evaluate the bispectrum
numerically and examine its configuration dependence and the effect of
the intrinsic source non-Gaussianity. We summarize our results and
discuss their implications in Sec.~V. The appendices provide additional
details of the calculations.

\section{PBH Evaporation and Induced Gravitational Waves}
\label{sec:model}

\subsection{Initial PBH Statistics and Isocurvature Source}
\label{subsec:poisson-source}

Let the comoving positions of PBHs be $\{\vect x_a\}$. We define the number density field as
\begin{equation}
 n(\vect x)=\sum_a\delD^{(3)}(\vect x-\vect x_a),
 \qquad
 S(\vect x)=\frac{n(\vect x)-\bar n_c}{\bar n_c},
 \label{eq:number-density-field}
\end{equation}
where $\bar n_c$ is the mean comoving number density. We work in the limit in which PBHs are spatially homogeneous on average, unclustered, and statistically independent, so that their distribution is described by an ideal Poisson process~\cite{DomenechLinSasaki2021,MoradinezhadDizgahFrancioliniRiotto2019}. For a homogeneous Poisson point process with intensity $\bar n_c$,
the probability generating functional of the random number-density measure $n(\vect x)$ is~\cite{LastPenrose2017}
\begin{equation}
	\ln Z_n[J]
	=
	\bar n_c
	\int\dd^3x\,
	\left[
	\ee^{J(\vect x)}-1
	\right].
	\label{eq:poisson-generating-functional}
\end{equation}
For any $N\geq2$ we adopt the coarse-grained description
\begin{align}
	\conavg{
		S(\vect k_1)\cdots S(\vect k_N)
	}
	={}&
	(2\pi)^3
	\delD^{(3)}
	\left(
	\sum_{a=1}^{N}\vect k_a
	\right)\nonumber\\
&\times
	\bar n_c^{\,1-N}
	W_n(\mathbf k_1,\ldots,\mathbf k_n),
	\label{eq:poisson-Npoint}
\end{align}
where $W_n(\mathbf k_1,\ldots,\mathbf k_n)\equiv
\prod_{i=1}^{n}\Theta(k_{\rm UV}-|\mathbf k_i|)$ is a window function, which implements the coarse-graining of the discrete distribution at the mean PBH separation scale:
\begin{equation}
 \kUV\equiv\frac{a_f}{d_f},
 \qquad
 d_f=\left(\frac{3M_{{\rm PBH},f}}{4\pi\rho_{{\rm PBH},f}}\right)^{1/3},
 \label{eq:kUV-def}
\end{equation}
where we adopt the monochromatic mass function $M_{{\rm PBH},f}=10^{4}\ {\rm g}$ for PBHs at formation. The mean comoving number density therefore satisfies
 \begin{equation} 
 	\bar n_c=a_f^3\bar n_f =a_f^3\frac{\rho_{{\rm PBH},f}}{M_{{\rm PBH},f}} =\frac{3}{4\pi}\kUV^3. \label{eq:nbarc-kUV}
\end{equation}
We define $ A_P\equiv\bar n_c^{-1}=\frac{4\pi}{3\kUV^3} \label{eq:AP-def} $; Eq.~\eqref{eq:poisson-Npoint} shows that the amplitude of the connected Poisson $N$-point function is proportional to $A_P^{N-1}$. In particular,
\begin{align}
 \avg{S(\vect k)S(\vect k')}
 &=(2\pi)^3\delD^{(3)}(\vect k+\vect k')P_S(k), \nonumber\\
 \langle S(\mathbf k_1)S(\mathbf k_2)S(\mathbf k_3)\rangle_c
 &=(2\pi)^3\delta_{\rm D}^{(3)}
 (\mathbf k_1+\mathbf k_2+\mathbf k_3)\nonumber\\
 &\quad\times B_S(\mathbf k_1,\mathbf k_2,\mathbf k_3).
 \label{eq:B_connected_def}
\end{align}
For the Poisson source, the white noise power spectrum and bispectrum are
\begin{align}
 P_S(k)
 &=A_PW_2(\mathbf k,-\mathbf k),
 \nonumber\\
 B_S(k_1,k_2,k_3)
 &={A_P}^2W_3(\mathbf k_1,\mathbf k_2,\mathbf k_3).
\end{align}
Their dimensionless counterparts are
 \begin{align}
 \mathcal P_S(k)
&\equiv \frac{k^3}{2\pi^2}P_S(k), \nonumber\\
 \mathcal B_S(k_1,k_2,k_3)
 &\equiv
 \left(\frac{\bar K^3}{2\pi^2}\right)^2
 B_S(k_1,k_2,k_3),
 \end{align}
where $\bar K\equiv \frac{k_1+k_2+k_3}{3}$.
 
We define the initial PBH fraction at formation as
\begin{equation}
	\beta\equiv
	\frac{\rho_{{\rm PBH},f}}{\rho_{{\rm tot},f}}
	\simeq
	\frac{\rho_{{\rm PBH},f}}{3\mpl^2H_f^2}
	,
	\qquad
	\rho_{{\rm PBH},f}=3\beta\mpl^2H_f^2,
	\label{eq:beta-def}
\end{equation}
where the cosmic Hubble parameter is $H\equiv\dot a/a$. We assume $\beta\ll1$, so the total energy density is still approximately dominated by radiation. 
We isolate the Poisson fluctuation of the PBH number density
as a pure initial isocurvature mode, leaving aside the independent
primordial adiabatic mode.
Its relative density perturbation with respect to radiation is
\begin{equation}
    S(\mathbf x,\tau)
    =
    \frac{\delta\rho_m}{\rho_m}
    -\frac34\frac{\delta\rho_r}{\rho_r}.
    \label{eq:isocurvature-def}
\end{equation}
For this mode, we take
\(\delta\rho_{r}=-\delta\rho_{m}\)
on the chosen initial slice. For monochromatic PBHs,
\(\delta\rho_{m}/\rho_{m}
=(n-\bar n_c)/\bar n_c\).
Since \(\rho_{m}\ll\rho_{r}\),
the radiation term in Eq.~\eqref{eq:isocurvature-def}
is subleading, so \(S_{\mathrm{in}}\simeq(n-\bar n_c)/\bar n_c\).
Thus the Poisson field defined in Eq.~\eqref{eq:number-density-field}
is the initial isocurvature source used below.
In the following, $S(\mathbf q)$ denotes the initial
stochastic field, while $S(\mathbf q,\tau)$ denotes its subsequent
time evolution.

\subsection{Evolution of Scalar Perturbations in the Two Fluid Background}
\label{subsec:scalar-evolution}
After PBH formation, the Universe can be treated as a two fluid system consisting of pressureless PBHs and radiation, with energy-momentum tensors
\begin{align}
	T_m^{\mu\nu}
	&=
	\rho_m u_m^\mu u_m^\nu,
	\\
	T_r^{\mu\nu}
	&=
	(\rho_r+p_r)u_r^\mu u_r^\nu
	+p_rg^{\mu\nu},
	\qquad
	p_r=\frac{\rho_r}{3}.
	\label{eq:two-fluid-stress-tensor}
\end{align}
We define the scalar velocity potentials $V_m$ and $V_r$ of the PBH and radiation fluids by
\begin{equation} u_{m\,i}=a\,\partial_iV_m, \qquad u_{r\,i}=a\,\partial_iV_r. \label{eq:velocity-potential-def} \end{equation} 
The difference $V_{\rm rel}\equiv V_m-V_r $ describes the relative velocity between the two components and is gauge invariant under first-order scalar gauge transformations.

The energy-momentum transfer between the two components due to PBH evaporation can be written as
\begin{equation}
	\nabla_\mu T_m^{\mu\nu}
	=
	-Q^\nu,
	\qquad
	\nabla_\mu T_r^{\mu\nu}
	=
	Q^\nu,
	\label{eq:energy-transfer}
\end{equation}
where
\begin{equation}
	Q^\nu=\Gamma\rho_m u_m^\nu,
	\qquad
	\Gamma
	\equiv
	-\frac{\dot M_{\rm PBH}}{M_{\rm PBH}}
	=
	-\frac{\dd\ln M_{\rm PBH}}{\dd t},
	\label{eq:Gamma-def}
\end{equation}
where a dot denotes a derivative with respect to cosmic time \(t\). Since $\Gamma\ll H$ during most of the PBH lifetime, we neglect all energy transfer between the components at early times and set $Q\simeq0,
\quad
\delta Q\simeq0$. Each component then redshifts independently. Because the radiation energy density decreases faster, a sufficiently large initial PBH abundance allows PBHs to dominate before they fully evaporate. At $\tau=\tau_{\rm eq}$, where $\rho_m=\rho_r$, the Universe transitions from early radiation-dominated (eRD) era to $\eMD$, and the background solution can be written compactly as
\begin{align}
	\frac{a(s)}{a_{\rm eq}}
	=A(s)\equiv\frac{\rho_m}{\rho_r}
	=\alpha s(\alpha s+2),\nonumber\\
	\text{where}
	s\equiv\frac{\tau}{\tau_{\rm eq}},
	\qquad
	\alpha\equiv\sqrt2-1.
	\label{eq:background-a-A}
\end{align}
For convenience, we introduce the dimensionless wavenumbers
\begin{equation}
	y\equiv\frac{q}{k_{\rm UV}},
	\qquad
	y_{\rm eq}\equiv\frac{k_{\rm eq}}{k_{\rm UV}},
	\label{eq:y-yeq-def}
\end{equation}
where \(k_{\rm eq}\equiv\mathcal H_{\rm eq}\) is the conformal Hubble scale. Throughout the numerical calculation, we fix
$
	y_{\rm eq}=4.308869\times10^{-2}.
$

For simplicity, we adopt the Newtonian gauge for the spacetime metric:
\begin{align}
 \dd s^2=a^2(\tau)\Bigl[&-(1+2\Psi)\dd\tau^2\nonumber\\
 &+\bigl((1+2\Phi)\delta_{ij}+h_{ij}\bigr)
 \dd x^i\dd x^j\Bigr].
 \label{eq:metric}
\end{align}
We also neglect first-order anisotropic stress. With the sign convention above, there is a single independent scalar perturbation, $\Phi=-\Psi$. We define
\begin{equation}
 \Phi(\vect q,\tau)=T_\Phi(q,\tau)S(\vect q),
 \qquad
 V_{\rm rel}(\vect q,\tau)=T_V(q,\tau)S(\vect q).
 \label{eq:transfer-def}
\end{equation}
Neglecting energy transfer, the first-order scalar dynamics through the eRD--eMD transition then obey
\begin{align}
 \Phi''&+3\Hc(1+c_s^2)\Phi'
 +\left[\Hc^2(1+3c_s^2)+2\Hc'-c_s^2\nabla^2\right]\Phi
 \nonumber\\
 &=\frac{a^2\rho_m c_s^2}{2\mpl^2}S,
 \label{eq:Phi-two-fluid}\\[0.3em]
 S'&=-\nabla^2V_{\rm rel},
 \label{eq:S-eom}\\
 V_{\rm rel}'
 &+3c_s^2\Hc V_{\rm rel}
 +\frac{3c_s^2\mpl^2}{2a^2\rho_r}\nabla^2\Phi
 +\frac{3\rho_m c_s^2}{4\rho_r}S
 =0,
 \label{eq:V-eom}
\end{align}
where \(c_s\) denotes the effective sound speed of the mixed fluid and satisfies
$c_s^2(s)
=
\frac{4}{9A(s)+12}$. Using $\Phi$, $\Pi=\dd\Phi/\dd s$, $S$, and $V_{\rm rel}$ as variables,
the scalar dynamics can be recast as a four-dimensional first-order
system convenient for numerical integration.
For each mode \(y\), we start the integration at the super‑horizon time
$
q\tau_{\rm in}=10^{-3}$
to ensure all modes have not entered the horizon initially.
The transfer functions are normalized to unit pure isocurvature perturbation, namely we take
\begin{equation}
	\Phi(\tau_{\rm in})=\Pi(\tau_{\rm in})=V_{\rm rel}(\tau_{\rm in})=0,
	\qquad
	S(\tau_{\rm in})=1.
	\label{eq:scalar-IC}
\end{equation}
These initial conditions determine the evolution of the scalar perturbations throughout the eRD--eMD era before the rapid final stage of PBH evaporation.

\subsection{Sudden Evaporation Transition}
\label{subsec:evap-matching}
After formation, PBHs continuously lose mass through Hawking radiation. If the spin-weighted effective number of degrees of freedom $g_{\rm H}(T_{\rm PBH})$ can be treated as approximately constant over the mass range of interest, the mass-loss rate can be written as
\begin{align}
	 \frac{\dd M_{\rm PBH}}{\dd t} &=-\frac{\alpha_{\rm H}M_{\rm Pl}^4}{M_{\rm PBH}^2}, \nonumber\\
	 \text{where }\alpha_{\rm H}&\equiv \frac{3.8\pi}{480}\,g_{\rm H}(T_{\rm PBH}), \qquad T_{\rm PBH}\equiv\frac{M_{\rm Pl}^2}{M_{\rm PBH}}. \label{eq:hawking-mass-loss} 
\end{align} 
We take $g_{\rm H}(T_{\rm PBH})=108$ for the calculation. Taking the PBH formation time $t_f$ as the origin of time and assuming $t_{\rm eva}\gg t_f$, integration gives
\begin{equation} M_{\rm PBH}(t) = M_{{\rm PBH},f} \left(1-\frac{t}{t_{\rm eva}}\right)^{1/3}, \qquad t_{\rm eva} = \frac{M_{{\rm PBH},f}^{3}} {3\alpha_{\rm H}M_{\rm Pl}^{4}} .
	\label{eq:PBH-mass-evolution}  
\end{equation}

If the initial PBH fraction is large enough for PBHs to dominate the radiation energy density before complete evaporation, so that $t_{\rm eva}\gg t_{\rm eq}>t_f$, the Universe passes from eRD into an approximately pure matter-dominated eMD era. Following~\cite{DomenechLinSasaki2021}, for a monochromatic PBH mass function and neglecting mass loss before the final evaporation stage, the condition for an eMD era is
$t_{\rm eq}<t_{\rm eva}$, equivalently
$\beta>\beta_{\rm dom}\equiv
\sqrt{\alpha_{\rm H}/(2\pi\gamma)}\,\mpl/M_{{\rm PBH},f}$.
To realize $t_{\rm eva}\gg t_{\rm eq}$ and hence a sufficiently long $\eMD$ era, one should take
$\beta\gg\beta_{\rm dom}$,
where $\gamma$ is the PBH formation efficiency defined by
$M_{{\rm PBH},f}=4\pi\gamma\mpl^2/H_f$
and is conventionally taken to be $\gamma\simeq0.2$ during radiation domination.

During the approximately matter-dominated stage, Eqs.~\eqref{eq:Gamma-def} and~\eqref{eq:PBH-mass-evolution} give
\begin{equation}
	\frac{\Gamma}{H}
	\simeq
	\frac12
	\left(
	\frac{t_{\rm eva}}{t}-1
	\right)^{-1}.
	\label{eq:Gamma-over-H}
\end{equation}
This ratio remains far below \(1\) during most of the PBH lifetime, but
rises rapidly as \(t\to t_{\rm eva}\). More specifically, defining \(\Gamma/H=1\) at $t_\star$ and using \(a\propto t^{2/3}\) during approximate matter domination, the expansion from
\(t_\star\) to complete evaporation is
\begin{equation}
	\Delta N_{\rm eva}
	\equiv
	\ln\frac{a(t_{\rm eva})}{a(t_\star)}\simeq0.27.
	\label{eq:last-evaporation-efold}
\end{equation}
Thus, during the final approximately $1/4$ e-fold, the evaporation rate exceeds the Hubble expansion rate and tends to diverge, so that energy is transferred rapidly into radiation near the end of the PBH lifetime. This motivates approximating the background evolution as
an instantaneous \(\eMD\to\lRD\) transition and idealizing $t_{eva}$ as an evaporation hypersurface ~\cite{DomenechLinSasaki2021}.

In the calculation, we stop the numerical evolution deep in the eMD era once the potential has reached its nearly constant plateau value $T_\Phi^{\rm eva}(y)$. We then extrapolate this value to the nominal evaporation time within the sudden-transition approximation. In the ideal instantaneous limit, evaporation takes place on an equal-proper-time hypersurface in the synchronous gauge comoving with PBHs. Transforming the junction conditions to the Newtonian gauge gives~\cite{DomenechLinSasaki2021}
\begin{equation}
[\Phi]_{\rm eva}=0,\qquad[\Phi']_{\rm eva}=0,
 \label{eq:Phi-matching}
\end{equation}
where the brackets denote the discontinuity across the evaporation hypersurface. Continuity of the background requires a shifted conformal time after evaporation,
\begin{equation}
 \bar\tau=\tau-\frac{\tau_{\rm eva}}{2},
 \qquad
 \Hc=\frac{1}{\bar\tau}.
 \label{eq:taubar}
\end{equation}

We define the comoving horizon scale at evaporation as
\begin{equation}
	k_{\rm eva}\equiv\mathcal H_{\rm eva}
	=\frac{1}{\bar\tau_{\rm eva}}
	=\frac{2}{\tau_{\rm eva}}.
\end{equation}
For a scalar mode $q$, we then define
$z=q\bar\tau$ and
$z_{\rm eva}=q\bar\tau_{\rm eva}=yR_{\rm eva}$,
where
\begin{equation}
R_{\rm eva}\equiv\frac{k_{\rm UV}}{k_{\rm eva}}
=2.3\times10^6.
\end{equation}

Together with the value of \(y_{\rm eq}\) specified above, \(R_{\rm eva}\) completes the choice of background scales used in the numerical calculation. For a monochromatic PBH population, these ratios can fix the initial PBH abundance and mass. Taking the conventional choices \(\gamma=0.2\) and \(g_H=108\), our benchmark corresponds approximately to
\[
\beta=1.2\times10^{-2},
\qquad
M_{{\rm PBH},f}=10^4\,{\rm g}.
\]
We use this fixed benchmark to illustrate the transfer evolution and the scale and configuration dependence of the tensor bispectrum.

Let $T_{\Phi,\lRD}^{(0)}$ denote the post-evaporation transfer function in the ideal instantaneous limit. Its general radiation-era solution, with $c_s=1/\sqrt3$, is

\begin{equation}
T_{\Phi,\lRD}^{(0)}(z;y)=\frac{1}{c_sz}
 \left[\mathcal A_\Phi(y)j_1(c_sz)+\mathcal B_\Phi(y)y_1(c_sz)\right].
 \label{eq:Tphi-lRD}
\end{equation}
The instantaneous junction conditions fix the coefficients $\mathcal A_\Phi(y)$ and $\mathcal B_\Phi(y)$ through
\begin{equation}
 T_{\Phi,\lRD}^{(0)}(z_{\rm eva};y)=T_\Phi^{\rm eva}(y),\qquad
 \left.\partial_zT_{\Phi,\lRD}^{(0)}(z;y)\right|_{z=z_{\rm eva}}\simeq0.
 \label{eq:lRD-matching}
\end{equation}
The instantaneous approximation captures the rapid background transition, but finite time evaporation reduces the amplitude of the post-evaporation scalar oscillations for modes with $q\gg k_{\rm eva}$. Radiation pressure reduces the efficiency with which PBH density perturbations are transferred to radiation perturbations and therefore suppresses the post-evaporation curvature potential. Following~\cite{Inomata2020,DomenechLinSasaki2021}, we account for this effect with the standard short-wavelength amplitude prescription
\begin{equation}
S_\Phi(y)\simeq\left(\sqrt{\frac23}\,yR_{\rm eva}\right)^{-1/3},\qquad yR_{\rm eva}\gg1.
 \label{eq:suppression-factor}
\end{equation}
For $yR_{\rm eva}\lesssim1$, we take $S_\Phi\simeq1$ and join the two regimes by a smooth interpolation in the numerical evaluation. The post-evaporation transfer function used below is the amplitude-corrected instantaneous solution,
\begin{align}
{T_{\Phi,\lRD}(z;y)}&\simeq S_\Phi(y)T_{\Phi,\lRD}^{(0)}(z;y),\nonumber\\
\widetilde T_\Phi^{\rm eva}(y)&\equiv S_\Phi(y)T_\Phi^{\rm eva}(y),
\end{align}
where $\widetilde T_\Phi^{\rm eva}$ specifies an effective radiation-era oscillation amplitude, rather than an additional exact junction condition on the instantaneous hypersurface. This amplitude prescription retains the oscillation phase of the instantaneous solution. For $z_{\rm eva}\gg1$, the leading terms are
\begin{align} T_{\Phi,\lRD}(z;y) &\simeq \widetilde T_\Phi^{\rm eva}(y) \left(\frac{z_{\rm eva}}{z}\right)^2 \cos\!\left[c_s(z-z_{\rm eva})\right], \label{eq:T-asym}\\ \partial_zT_{\Phi,\lRD}(z;y) &\simeq -c_s\widetilde T_\Phi^{\rm eva}(y) \left(\frac{z_{\rm eva}}{z}\right)^2 \sin\!\left[c_s(z-z_{\rm eva})\right]. \label{eq:dT-asym} 
\end{align} 

Although $\Phi'\simeq0$ at the ideal matching surface, radiation pressure drives acoustic oscillations over a time scale $\Delta\bar\tau\sim(c_sq)^{-1}\ll\mathcal H_{\rm eva}^{-1}$. The resulting large derivative of the potential provides the enhanced post-evaporation source of induced GWs~\cite{DomenechLinSasaki2021}.

\subsection{Second-Order Tensor Response}
\label{subsec:tensor-kernel}

We decompose the tensor perturbation $h_{ij}$ into two polarization modes,
\begin{equation}
 h_{ij}(\tau,\vect x)=\sum_{\lambda}\intk
 e_{ij}^{\lambda}(\hat{\vect k})h_{\vect k}^{\lambda}(\tau)
 \ee^{\ii\vect k\cdot\vect x},
\end{equation}
where $\lambda=R,L$, with $R$ and $L$ denoting right- and left-handed circular polarizations, respectively. Each polarization mode obeys
\begin{equation}
 h_{\vect k}^{\lambda\prime\prime}+2\Hc h_{\vect k}^{\lambda\prime}+k^2h_{\vect k}^{\lambda}
 =\mathcal S_{\vect k}^{\lambda}.
 \label{eq:tensor-eom}
\end{equation}
For the PBH--radiation two-fluid system, the scalar source at second order can be written as
\begin{align}
	\mathcal S_{ij}={}&
	4\partial_i\Phi\partial_j\Phi
	+6c_s^2\frac{\rho}{\rho_r}
	\partial_i\!\left(\frac{\Phi'}{\Hc}+\Phi\right)
	\partial_j\!\left(\frac{\Phi'}{\Hc}+\Phi\right)
	\nonumber\\
	&+\frac{6a^2c_s^2\rho_m}{\mpl^2}
	\partial_iV_{\rm rel}\partial_jV_{\rm rel},
	\label{eq:tensor-source}
\end{align}
where $\rho\equiv\rho_m+\rho_r$. In Fourier space, this becomes
\begin{align}
 \mathcal S_{\mathbf k}^{\lambda}(\tau)
 ={}&-4k^2
 \int\frac{\dd^3p}{(2\pi)^3}\,
 \Pi_\lambda(\mathbf k,\mathbf p)\,
 \mathcal F(u,v;k,\tau)
 \nonumber\\
 &\times S(\mathbf p)S(\mathbf k-\mathbf p),
 \label{eq:source-F-def}
\end{align}
where
\begin{align}
 \mathcal F(u,v;k,\tau)
 ={}&T_\Phi(vk,\tau)T_\Phi(uk,\tau)
 \nonumber\\
 &+\frac{3\rho}{2\rho_r}c_s^2
 \left[
 \frac{T_\Phi'(vk,\tau)}{\Hc}+T_\Phi(vk,\tau)
 \right]
 \nonumber\\
 &\quad\times
 \left[
 \frac{T_\Phi'(uk,\tau)}{\Hc}+T_\Phi(uk,\tau)
 \right]
 \nonumber\\
 &+\frac{3a^2c_s^2\rho_m}{2\mpl^2}
 T_V(vk,\tau)T_V(uk,\tau),
 \label{eq:F-full}
\end{align}
with $u=\frac{|\vect k-\vect p|}{k}$ and
$v=\frac{p}{k}$. 

We define the Green function by
\begin{equation} 
	\left[
	\partial_\tau^2+k^2-\frac{a''(\tau)}{a(\tau)}
	\right]
	G_k(\tau,\tau')
	=
	\delD(\tau-\tau'). \label{eq:Green-def}
\end{equation} 
The formal solution for the tensor mode is
\begin{align}
 h_{\mathbf k}^{\lambda}(\tau_{\rm out})
 ={}&-4
 \int\frac{\dd^3p}{(2\pi)^3}\,
 \Pi_\lambda(\mathbf k,\mathbf p)\,
 I(u,v;k,\tau_{\rm out})
 \nonumber\\
 &\times S(\mathbf p)S(\mathbf k-\mathbf p).
 \label{eq:h-kernel-compact}
\end{align}
We have introduced the polarization contraction
\begin{equation}
	\Pi_\lambda(\mathbf k,\mathbf p)
	\equiv
	e^\lambda_{ij}(\hat{\mathbf k})
	\frac{p^ip^j}{k^2}.
	\label{eq:Pi-def}
\end{equation}

The dimensionless time kernel is defined as
\begin{align}
 I(u,v;k,\tau_{\rm out})
 \equiv{}&
 \frac{k^2}{a(\tau_{\rm out})}
 \int_{\tau_{\rm in}}^{\tau_{\rm out}}
 \dd\tau'\,
 G_k(\tau_{\rm out},\tau')
 \nonumber\\
 &\times a(\tau')\,
 \mathcal F(u,v;k,\tau'),
 \label{eq:I-general-def}
\end{align}
where we choose the late‑time observational surface as $
k_{\rm UV}\bar\tau_{\rm out}=1.5R_{\rm eva}
=3.45\times10^6.$
Thus, the source statistics and polarization geometry are retained separately in $S(\vect p)S(\vect k-\vect p)$ and $\Pi_\lambda$, while the background evolution and the transfer functions of the two scalar source legs are packaged into $I$.

In principle,  $I=I_{\eRD}+I_{\eMD}+I_{\lRD}$. 
Following~\cite{DomenechLinSasaki2021}, we retain only the dominant post-evaporation contribution to \(I_{\rm lRD}\), arising from the term quadratic in the time derivatives of the scalar potential in~\eqref{eq:F-full}. Deep in the eMD era, the curvature potential is approximately frozen, whereas the sudden transition excites rapid acoustic oscillations immediately after evaporation. For modes well inside the horizon at evaporation, their characteristic amplitudes satisfy
\begin{equation}
	\left|\frac{\Phi'}{\mathcal H}\right|
	\sim
	c_s z_{\rm eva}|\Phi|
	\gg |\Phi|.
\end{equation}
Consequently, the term proportional to $(\Phi'/\mathcal{H})$ dominates the enhanced post-evaporation source, while the non-derivative terms give subleading contributions to the sudden-transition enhancement. We therefore approximate the time kernel by this dominant lRD contribution.
To organize the late time kernel, we first define
\begin{align}
 K&\equiv\frac{k}{k_{\rm UV}},
 \nonumber\\
 x_{\rm out}&\equiv k\bar\tau_{\rm out},
 \nonumber\\
 x_{\rm eva}&\equiv k\tau_{\rm eva}=2KR_{\rm eva},
 \nonumber\\
 x_3&\equiv k\bar\tau_{\rm eva}=\frac{x_{\rm eva}}{2},
 \nonumber\\
 x_2&\equiv
 k\left(\bar\tau-\bar\tau_{\rm eva}\right),
 \nonumber\\
 x_1&\equiv
 k\left(\bar\tau_{\rm out}-\bar\tau_{\rm eva}\right)
 =x_{\rm out}-x_3,
 \nonumber
\end{align}
where \(x_2\) is the integration variable measured from the evaporation hypersurface and satisfies $x_2+x_3=k\bar\tau$. The dimensionless wavenumbers of the two internal scalar modes are
\begin{equation}
	y_u\equiv\frac{uk}{k_{\rm UV}}=uK,
	\qquad
	y_v\equiv\frac{vk}{k_{\rm UV}}=vK.
	\label{eq:yu-yv-def}
\end{equation}

Under the approximations above, the dimensionless time kernel can finally be written as
\begin{align}
	I_{\lRD}(u,v;K,x_{\rm out})
	\simeq{}&
	\frac{c_s^2uv\,x_{\rm eva}^4}{32x_{\rm out}}\,
	\widetilde T_\Phi^{\rm eva}(y_u)\,
	\widetilde T_\Phi^{\rm eva}(y_v)
	\nonumber\\
	&\times
	\mathcal J(c_su,c_sv;x_1,x_3),
	\label{eq:I-lRD}
\end{align}
where
\begin{align}
 \mathcal J(a,b;x_1,x_3)
 \equiv{}&
 \int_0^{x_1}
 \frac{\dd x_2}{x_2+x_3}\,
 \sin(x_1-x_2)
 \nonumber\\
 &\times\sin(ax_2)\sin(bx_2).
 \label{eq:J-kernel-def}
\end{align}
For the observational surface adopted below,
$x_1\gg1$ throughout the numerical domain. Away from resonance, the integral in Eq.~\eqref{eq:J-kernel-def} oscillates with $x_1$, while its oscillation amplitude approaches a constant in the limit $x_1\rightarrow \infty$. We therefore take the late-time limit and extend the upper limit of the
time integral to infinity, following~\cite{DomenechLinSasaki2021}. The condition $a+b=1$ corresponds to the resonance, at which the resulting asymptotic kernel develops a logarithmic singularity. This singularity is integrable in the subsequent momentum integration and therefore does not lead to a divergence of the tensor correlators. We discuss the analytic structure of the integral and the property of the resonance band as detailed in Appendix~\ref{app:lRD-kernel}.

\section{Selection Rules for the Tensor Spectrum Decomposition}
\label{sec:formalism}
 Equation~\eqref{eq:h-kernel-compact} writes each tensor mode as a quadratic convolution of the scalar isocurvature source. A tensor \(n\)-point function therefore corresponds to a source-field \(S\) correlator of order \(2n\). Since a Poisson source has nonzero connected correlation functions at all orders, the expansion of the tensor \(n\)-point function contains not only Wick contracted contributions but also combinations of higher-order connected correlators, for example
\begin{align}
	\langle h h\rangle
	&\sim
	\langle S^4\rangle
	\sim
	PP+T ,\\
	\langle h h h\rangle
	&\sim
	\langle S^6\rangle
	\sim
	PPP+BB+PT+C_6 ,
\end{align}
where we use $P,B,T$ and $C_6$ to label connected source blocks containing two, three, four, and six source fields; products such
as $PP$ and $PT$ include the corresponding partitions of the
source legs.
At the level of this expansion, the different topological combinations of source correlators have the same parametric scaling. To see this, note from Eq.~\eqref{eq:poisson-Npoint} that the amplitude of a connected $N$-point block of an ideal Poisson source scales as
\begin{equation}
 A_P^{N-1}\sim\kUV^{-3(N-1)}.
\end{equation}
Consider an $m$-point tensor correlator, which contains $2m$ scalar source legs. If a given correlator decomposes into $r$ connected blocks of sizes $n_1,\ldots,n_r$, with $\sum_{a=1}^r n_a=2m$, its Poisson amplitude is \begin{equation} \prod_{a=1}^{r}A_P^{n_a-1} =A_P^{2m-r}. \end{equation} The $r$ connected blocks supply $r$ momentum-conserving delta functions. After factoring out the overall external momentum-conservation delta function and eliminating the internal integrals, a naive dimensional estimate, apart from the common integral kernel, gives \begin{equation} A_P^{2m-r} \left(\kUV^3\right)^{m-r+1} \sim \kUV^{-3(m-1)}, \label{eq:poisson-topology-power-counting} \end{equation} 
which is independent of the number of connected blocks $r$. Therefore, the contribution of $PP$ and $T$ terms in the tensor power spectrum, and the $PPP$, $BB$, $PT$, and $C_6$ terms in the bispectrum, cannot be excluded in the sense of scale parameter before the tensor projection is imposed.

We next further determine which contractions can give nonzero contributions under TT projection by representing the connected decomposition of the source field as  vertex‑connected diagrams. Each tensor mode $h_{\mathbf k_i}^{\lambda_i}$ is represented by a quadratic source vertex: one external tensor leg $h_{\mathbf k_i}^{\lambda_i}$ is attached to two scalar source legs $\bigl(S(\mathbf p_i),S(\mathbf q_i)\bigr)$. 
Each connected source $n$-point function is represented by
a vertex joining $n$ scalar source lines. The topological selection rule is then as follows: among all contractions in the connected-function expansion, any contraction that joins the two source points belonging to the same tensor vertex---forming a ``self-loop'' on that vertex---vanishes. Such a term contains a separable single-leg integral in which the kernel $F$, apart from the polarization projection, is axisymmetric around the external tensor momentum $\mathbf k$, namely
\begin{align} \mathscr L_\lambda(\mathbf k) &\equiv \int\dd^3p\, \Pi_\lambda(\mathbf k,\mathbf p)\, F\!\left(p,\hat{\mathbf k}\cdot\hat{\mathbf p}\right)\nonumber\\
	&=e^{\lambda}_{ij}(\hat{\mathbf k})(A\delta_{ij}+B\hat k_i\hat k_j).\end{align} 
This integral vanishes by
$e^{\lambda}_{ii}=0$ and
$e^{\lambda}_{ij}\hat k_j=0$.
Consequently, only contractions without a separable
axisymmetric single-leg integral can survive the TT projection.

	\begin{figure}[t]
		\centering
		\resizebox{\columnwidth}{!}{%
			\begin{tikzpicture}[
				x=1.15cm,
				y=1cm,
				field/.style={
					inner sep=1.2pt,
					outer sep=0pt,
					font=\normalsize
				},
				leg/.style={
					font=\scriptsize,
					text=gray!70!black
				},
				conn/.style={
					line width=0.4pt
				},
				result/.style={
					font=\normalsize,
					anchor=west
				}
				]
				
				\node[font=\normalsize,anchor=east]
				at (-0.70,0) {(a) $PPP$};
				
				\node at (-0.42,0) {$\langle$};
				
				\node[field] (ppp1) at (0,0)    {$S_{1p}$};
				\node[field] (ppp2) at (1,0)    {$S_{1q}$};
				\node[field] (ppp3) at (2.35,0) {$S_{2p}$};
				\node[field] (ppp4) at (3.35,0) {$S_{2q}$};
				\node[field] (ppp5) at (4.70,0) {$S_{3p}$};
				\node[field] (ppp6) at (5.70,0) {$S_{3q}$};
				
				\node at (6.12,0) {$\rangle$};
				\node[result] at (6.50,0) {$\neq0$};
				
				\node[leg] at (0.50,0.38) {$h_1$};
				\node[leg] at (2.85,0.38) {$h_2$};
				\node[leg] at (5.20,0.38) {$h_3$};
				
				\draw[conn]
				(ppp2.south)
				-- ($(ppp2.south)+(0,-0.35)$)
				-- ($(ppp3.south)+(0,-0.35)$)
				-- (ppp3.south);
				
				\draw[conn]
				(ppp4.south)
				-- ($(ppp4.south)+(0,-0.35)$)
				-- ($(ppp5.south)+(0,-0.35)$)
				-- (ppp5.south);
				
				\draw[conn]
				(ppp1.south)
				-- ($(ppp1.south)+(0,-0.72)$)
				-- ($(ppp6.south)+(0,-0.72)$)
				-- (ppp6.south);
				
				\node[font=\scriptsize,fill=white,inner sep=0.5pt]
				at (1.675,-0.35) {$P$};
				\node[font=\scriptsize,fill=white,inner sep=0.5pt]
				at (4.025,-0.35) {$P$};
				\node[font=\scriptsize,fill=white,inner sep=0.5pt]
				at (2.85,-0.72) {$P$};

				\node[font=\normalsize,anchor=east]
				at (-0.70,-1.80) {(b) $BB$};
				
				\node at (-0.42,-1.80) {$\langle$};
				
				\node[field] (bb1) at (0,-1.80)    {$S_{1p}$};
				\node[field] (bb2) at (1,-1.80)    {$S_{1q}$};
				\node[field] (bb3) at (2.35,-1.80) {$S_{2p}$};
				\node[field] (bb4) at (3.35,-1.80) {$S_{2q}$};
				\node[field] (bb5) at (4.70,-1.80) {$S_{3p}$};
				\node[field] (bb6) at (5.70,-1.80) {$S_{3q}$};
				
				\node at (6.12,-1.80) {$\rangle$};
				\node[result] at (6.50,-1.80) {$\neq0$};
				
				\node[leg] at (0.50,-1.42) {$h_1$};
				\node[leg] at (2.85,-1.42) {$h_2$};
				\node[leg] at (5.20,-1.42) {$h_3$};
				
				\coordinate (bbA) at (0,-2.28);
				
				\draw[conn] (bb1.south) -- (bb1.south |- bbA);
				\draw[conn] (bb3.south) -- (bb3.south |- bbA);
				\draw[conn] (bb5.south) -- (bb5.south |- bbA);
				\draw[conn]
				(bb1.south |- bbA)
				-- (bb5.south |- bbA);
				
				\node[font=\scriptsize,fill=white,inner sep=0.5pt]
				at (2.35,-2.28) {$B$};
				
				\coordinate (bbB) at (0,-2.68);
				
				\draw[conn] (bb2.south) -- (bb2.south |- bbB);
				\draw[conn] (bb4.south) -- (bb4.south |- bbB);
				\draw[conn] (bb6.south) -- (bb6.south |- bbB);
				\draw[conn]
				(bb2.south |- bbB)
				-- (bb6.south |- bbB);
				
				\node[font=\scriptsize,fill=white,inner sep=0.5pt]
				at (3.35,-2.68) {$B$};

				\node[font=\normalsize,anchor=east]
				at (-0.70,-3.80) {(c) $PT$};
				
				\node at (-0.42,-3.80) {$\langle$};
				
				\node[field] (pt1) at (0,-3.80)    {$S_{1p}$};
				\node[field] (pt2) at (1,-3.80)    {$S_{1q}$};
				\node[field] (pt3) at (2.35,-3.80) {$S_{2p}$};
				\node[field] (pt4) at (3.35,-3.80) {$S_{2q}$};
				\node[field] (pt5) at (4.70,-3.80) {$S_{3p}$};
				\node[field] (pt6) at (5.70,-3.80) {$S_{3q}$};
				
				\node at (6.12,-3.80) {$\rangle$};
				\node[result] at (6.50,-3.80) {$=0$};
				
				\node[leg] at (0.50,-3.42) {$h_1$};
				\node[leg] at (2.85,-3.42) {$h_2$};
				\node[leg] at (5.20,-3.42) {$h_3$};
				
				\draw[conn]
				(pt1.south)
				-- ($(pt1.south)+(0,-0.38)$)
				-- ($(pt3.south)+(0,-0.38)$)
				-- (pt3.south);
				
				\node[font=\scriptsize,fill=white,inner sep=0.5pt]
				at (1.175,-4.18) {$P$};
				
				\coordinate (ptT) at (0,-4.62);
				
				\foreach \n in {pt2,pt4,pt5,pt6}{
					\draw[conn] (\n.south) -- (\n.south |- ptT);
				}
				
				\draw[conn]
				(pt2.south |- ptT)
				-- (pt6.south |- ptT);
				
				\node[font=\scriptsize,fill=white,inner sep=0.5pt]
				at (3.35,-4.62) {$T$};
				
				\node[font=\scriptsize,anchor=west]
				at (4.25,-5.02)
				{$\mathscr L_t(\mathbf k_3)=0$};

				\node[font=\normalsize,anchor=east]
				at (-0.70,-5.95) {(d) $C_6$};
				
				\node at (-0.42,-5.95) {$\langle$};
				
				\node[field] (c61) at (0,-5.95)    {$S_{1p}$};
				\node[field] (c62) at (1,-5.95)    {$S_{1q}$};
				\node[field] (c63) at (2.35,-5.95) {$S_{2p}$};
				\node[field] (c64) at (3.35,-5.95) {$S_{2q}$};
				\node[field] (c65) at (4.70,-5.95) {$S_{3p}$};
				\node[field] (c66) at (5.70,-5.95) {$S_{3q}$};
				
				\node at (6.12,-5.95) {$\rangle_c$};
				\node[result] at (6.65,-5.95) {$=0$};
				
				\node[leg] at (0.50,-5.57) {$h_1$};
				\node[leg] at (2.85,-5.57) {$h_2$};
				\node[leg] at (5.20,-5.57) {$h_3$};
				
				\coordinate (c6L) at (0,-6.48);
				
				\foreach \n in {c61,c62,c63,c64,c65,c66}{
					\draw[conn] (\n.south) -- (\n.south |- c6L);
				}
				
				\draw[conn]
				(c61.south |- c6L)
				-- (c66.south |- c6L);
				
				\node[font=\scriptsize,fill=white,inner sep=0.5pt]
				at (2.85,-6.48) {$C_6$};
				
				\node[font=\scriptsize]
				at (2.85,-6.88)
				{$
					\mathscr L_r(\mathbf k_1)
					\mathscr L_s(\mathbf k_2)
					\mathscr L_t(\mathbf k_3)=0
					$};
				
			\end{tikzpicture}%
		}
		\caption{
			Representative contractions for the four topological classes in the decomposition of the source six-point moment entering the tensor bispectrum. The first subscript of S labels the external tensor leg to which the corresponding source field belongs.
			(a) A nontrivial cyclic $PPP$ contraction;
			(b) a nonzero $BB$ contraction in which each connected three-point block contains one source field from each external leg;
			(c) every $PT$ contraction forms a ``self loop'' and vanishes;
			(d) the fully connected six-point term leaves only overall external momentum conservation and vanishes.
		}
		\label{fig:sixpoint-contractions}
	\end{figure}
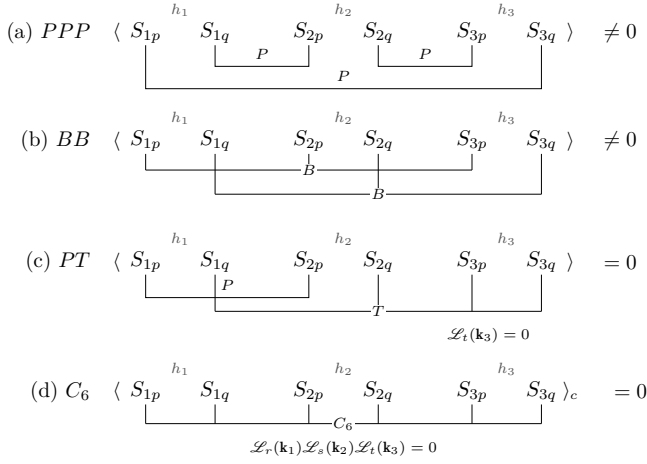

Figure~\ref{fig:sixpoint-contractions}
shows four representative connected topologies in the source six-point function entering the tensor bispectrum. The selection rule above reduces the nontrivial six-point decomposition to
\begin{equation}
	\langle hhh\rangle \sim\langle SSSSSS\rangle
	\sim
	PPP+BB.
\end{equation}
Among the 15 equivalent contractions of the \(PPP\) type, 8 contribute, while only 4 of the 10 contractions built from two connected three-point functions of the \(BB\) type contribute. Similarly, for the source four-point function entering the power spectrum, only the $PP$ contractions are nontrivial, whereas the connected $T$ contribution vanishes.

The analysis above makes explicit how higher-order source statistics enter the successive spectra of the induced GWs. It also shows the additional information carried by the tensor bispectrum: higher-order source cumulants do not enter the tensor power spectrum. Consequently, an ideal Poisson source and a Gaussianized source with the same white noise
two-point spectrum are degenerate at the level of the tensor power spectrum. By contrast, through the \(BB\) contribution, the tensor bispectrum retains higher-order non-Gaussian information about source discreteness that is invisible in the power spectrum. 

This conclusion can be generalized beyond a monochromatic PBH point process. Consider a statistically homogeneous source \(X\),
whose connected correlation functions satisfy
\begin{align}
 \left\langle
 X(\mathbf k_1)\cdots X(\mathbf k_N)
 \right\rangle_c
 ={}&(2\pi)^3
 \delta_{\rm D}^{(3)}
 \left(
 \sum_{a=1}^{N}\mathbf k_a
 \right)
 \nonumber\\
 &\times\mathcal C_N(k_1,\ldots,k_N).
 \label{eq:shot-noise-connected-class}
\end{align}
If the tensor mode can still be written as the quadratic convolution in Eq.~\eqref{eq:h-kernel-compact},
if the time kernel introduces no preferred direction for the external momentum, and if
\(\mathcal C_N\) depends only on the magnitudes of the momenta and contains no mixed angular dependence, the same selection rule continues to hold.

The tensor three-point function is finally defined by
\begin{equation}
 \avg{h^r_{\vect k_1}h^s_{\vect k_2}h^t_{\vect k_3}}
 =(2\pi)^3\delD^{(3)}(\vect k_1+\vect k_2+\vect k_3)
 B_h^{rst}(k_1,k_2,k_3),
 \label{eq:Bh-def}
\end{equation}
where \(B_h^{rst}\), the dimensionful reduced tensor bispectrum, is decomposed as
\begin{equation}
B_h^{rst}(k_1,k_2,k_3)
=
B_{h,PPP}^{rst}(k_1,k_2,k_3)
+
B_{h,BB}^{rst}(k_1,k_2,k_3).
 \label{eq:Bh-selection-final}
\end{equation}
The $PPP$ class contains eight equivalent nontrivial contractions, whose total contribution is
\begin{align}
	B_{h,PPP}^{rst}
	={}&
	-8\times64
	\int\frac{\dd^3p}{(2\pi)^3}\,
	\Pi_1^r\Pi_2^s\Pi_3^t
	\nonumber\\
	&\times
	P_S(p_1)P_S(p_2)P_S(p_3)
	I_1I_2I_3,
	\label{eq:Bh-PPP-dimensionful}
\end{align}
where $\Pi_i^r\equiv\Pi_r(\mathbf k_i,\mathbf p_i)$, $I_i\equiv I(u_i,v_i;k_i,\tau_{\rm out})$, $
	\vect p_1=\vect p,
	\vect p_2=\vect p-\vect k_1,
	\vect p_3=\vect p+\vect k_3$. For each tensor vertex, we define 
$
	\mathbf q_i\equiv\mathbf k_i-\mathbf p_i,
	\quad i=1,2,3$.

The $BB$ class contains four equivalent nontrivial contractions, with the total contribution
\begin{align}
	B_{h,BB}^{rst}
	={}&
	-4\times64
	\int\frac{\dd^3p_1}{(2\pi)^3}
	\frac{\dd^3p_2}{(2\pi)^3}\,
	\Pi_1^r\Pi_2^s\Pi_3^t
	I_1I_2I_3
	\nonumber\\
	&\times
	B_S(\mathbf p_1,\mathbf p_2,\mathbf p_3)
	B_S(\mathbf q_1,\mathbf q_2,\mathbf q_3)
	\bigg|_{\mathbf p_3=-\mathbf p_1-\mathbf p_2}.
	\label{eq:Bh-BB-dimensionful}
\end{align}

 To match the numerical implementation directly, we further define the dimensionless bispectrum
\begin{equation}
	\widehat{\mathcal B}_h^{rst}
	\equiv
	(k_1k_2k_3)^2
	B_h^{rst}(k_1,k_2,k_3).
	\label{eq:dimensionless-Bh-code}
\end{equation}
The numerical figures use the bispectrum kernel normalized by $N_P
=
-\frac{512\,\widehat A_P^{\,3}}{(2\pi)^3}
$, where the dimensionless Poisson amplitude is $
	\widehat A_P
	\equiv k_{\rm UV}^3A_P
	=\frac{4\pi}{3}$.
\begin{align}
	\mathcal B_{\rm tot}^{rst}
	\equiv
	\frac{\widehat{\mathcal B}_h^{rst}}{N_P}\equiv
	\mathcal B_{PPP}^{rst}
	+
	\mathcal B_{BB}^{rst}.
	\label{eq:Btot-prefactor-convention}
\end{align}

Also, in numerical implementation we characterize each oscillatory time kernel $I$ by its envelope 
\begin{equation}
	I_{\rm env}(u,v;k)=\left[\frac{I_s^2(u,v;k)+I_c^2(u,v;k)}{2}\right]^{1/2},
	\label{eq:Ienv}
\end{equation}
where $I_s$ and $I_c$ are defined by the decomposition 
\begin{equation}
	I(u,v;k)
	=
	I_s(u,v;k)\sin x_{\rm out}
	+
	I_c(u,v;k)\cos x_{\rm out}.
\end{equation}


\section{Numerical Results and Physical Interpretation}
\label{sec:numerical-results}

\begin{figure}[t]
	\centering
	\begin{subfigure}{0.9\columnwidth}
		\centering
		\includegraphics[width=\linewidth]{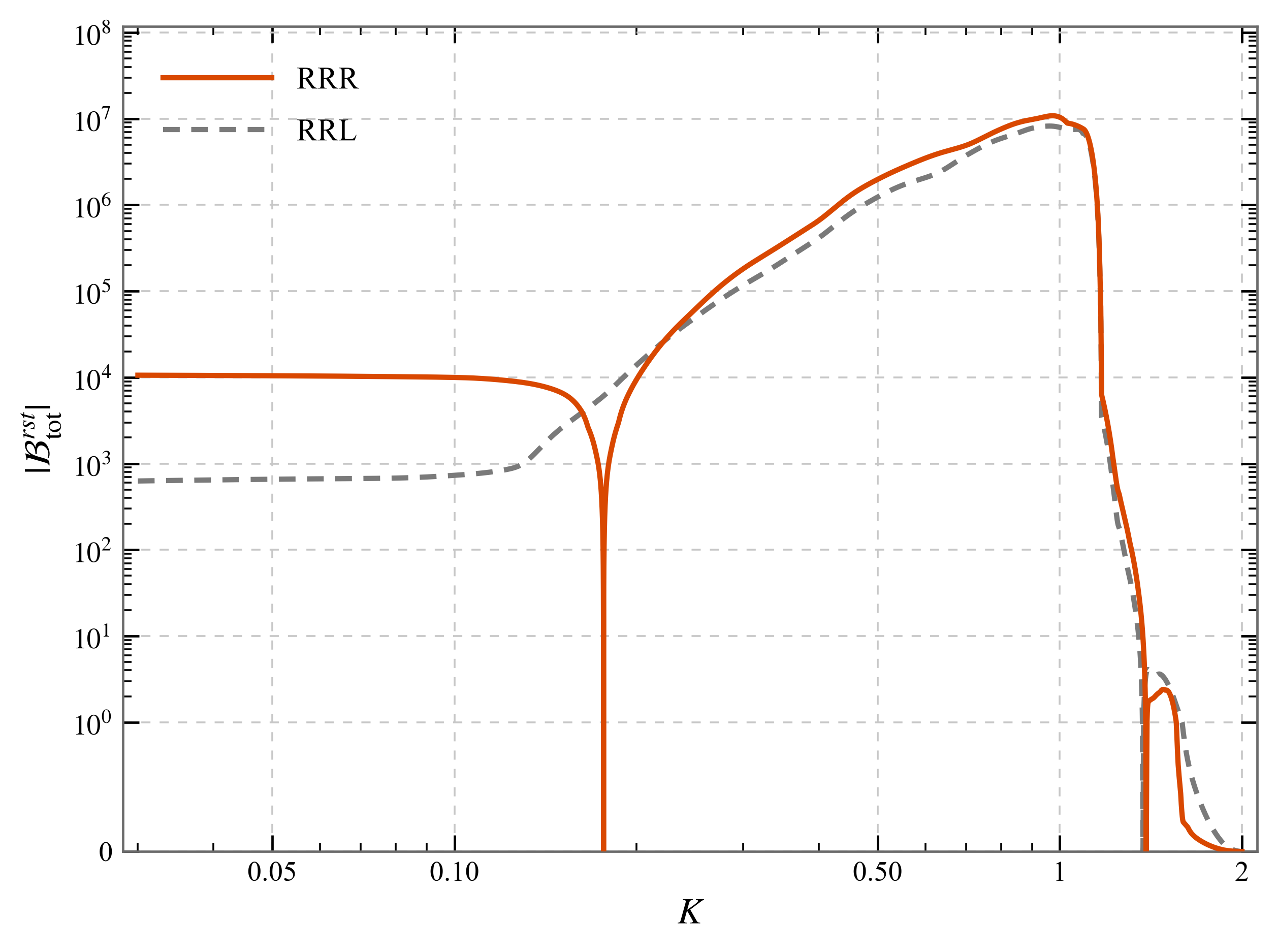}
		\caption{}  
		\label{fig:equilateral-polarizations}
	\end{subfigure}
	
	\vspace{0.6em}  
	
	\begin{subfigure}{0.9\columnwidth}
		\centering
		\includegraphics[width=\linewidth]{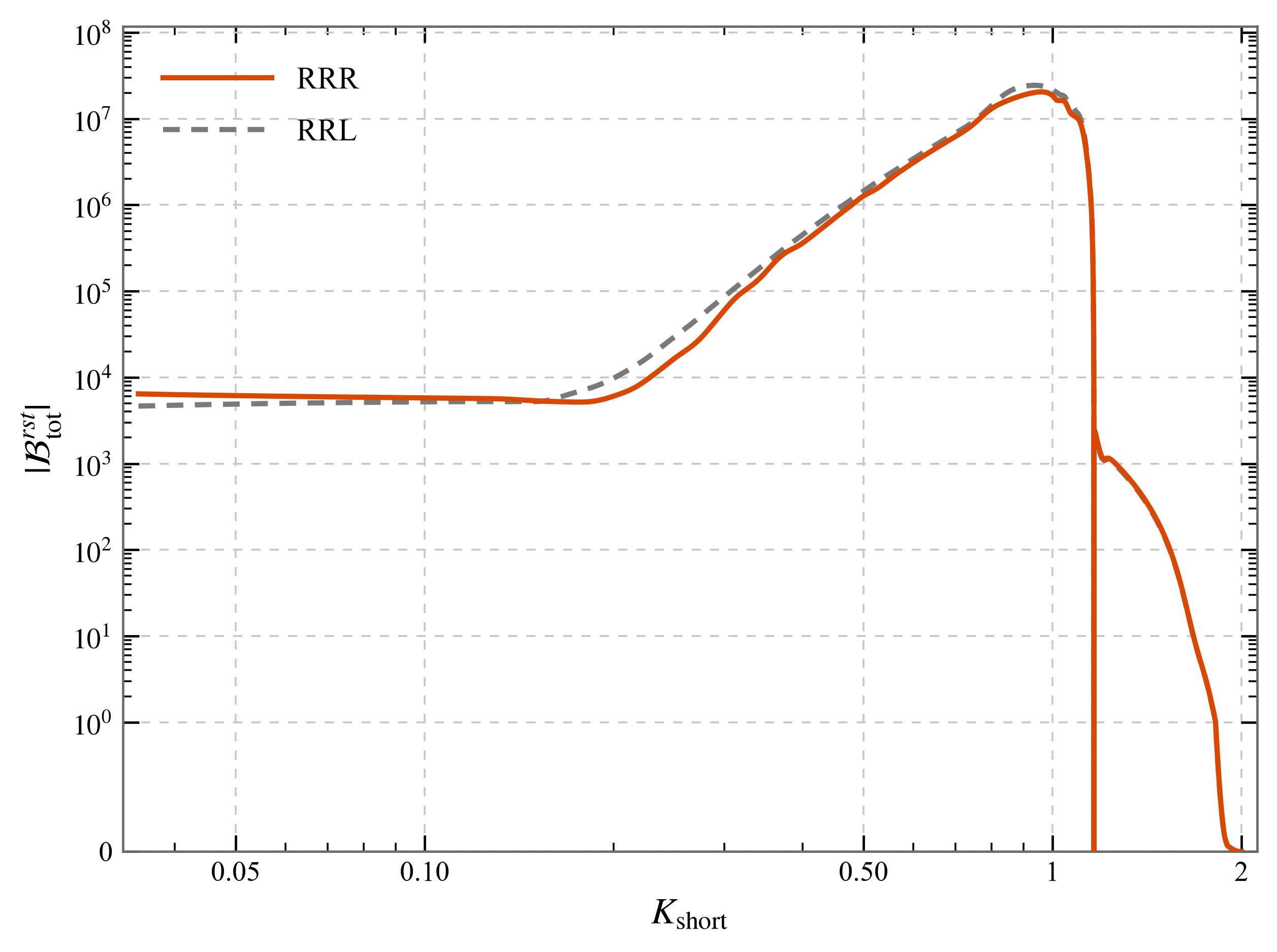}
		\caption{}  
		\label{fig:squeezed-polarizations}
	\end{subfigure}
	
	\caption{
		Absolute value of the total bispectrum of GWs induced by a Poisson isocurvature source through sudden PBH evaporation. The red solid and gray dashed curves denote the $RRR$ and $RRL$ polarization channels, respectively.
		(a) variation in the equilateral configuration $K_1=K_2=K_3=K$ with the side length $K$;
		(b) variation in the squeezed configuration $K_1=K_2=K_{\rm short}$ and $K_3=0.005$ .
	}
	\label{fig:polarizations_both}  
\end{figure}

In this section, we numerically evaluate the bispectrum of GWs induced by a Poisson isocurvature source through sudden PBH evaporation. We focus on the new information carried by the higher-order connected $BB$ contribution in the total bispectrum, beyond the Wick $PPP$ contribution. For compactness, all momenta in this section are expressed in dimensionless form:
\begin{equation}
	\bar{\mathbf k}\equiv \frac{\mathbf k}{k_{\rm UV}},
	\qquad
	K\equiv |\bar{\mathbf k}|.
	\label{eq:dimensionless_momenta_for_spectra}
\end{equation}
For the external momentum triangle, we further define the dimensionless perimeter
\begin{equation}
	\bar C\equiv K_1+K_2+K_3
	=\frac{k_1+k_2+k_3}{k_{\rm UV}} .
	\label{eq:Cbar_def}
\end{equation}

We first present the overall bispectrum shapes for different polarizations in the equilateral and squeezed configurations. Parity symmetry in the equilateral configuration leaves only two independent polarization combinations, $\mathrm{RRR}$ and $\mathrm{RRL}$, as shown in Fig.~\ref{fig:polarizations_both}(\subref{fig:equilateral-polarizations}). For $K\lesssim0.15$,
both channels approach plateaus with the amplitude of $RRR$ larger than $RRL$ by a factor of order unity but opposite signs. The $RRR$ channel crosses zero near
$K\simeq0.18$,
changing from negative to positive, while $RRL$ is always positive. Thereafter, both polarization channels grow with wavenumber and reach approximate maxima around $K\sim 1$, nearly overlapping with each other. Between K=1 and K=2, each curve changes sign, reaches a smaller local extremum, and vanishes at K=2. As in the power spectrum result of~\cite{DomenechLinSasaki2021}, a peak near the cutoff scale is expected because the source correlators and the derivatives of the transfer functions contribute factors of the form $k^n$, whereas for $K>1$ the shrinking effective integration domain imposed by the cutoff suppresses the spectrum and eventually drives it to zero.

\begin{figure}[t]
	\centering
	\includegraphics[width=0.9\columnwidth]
	{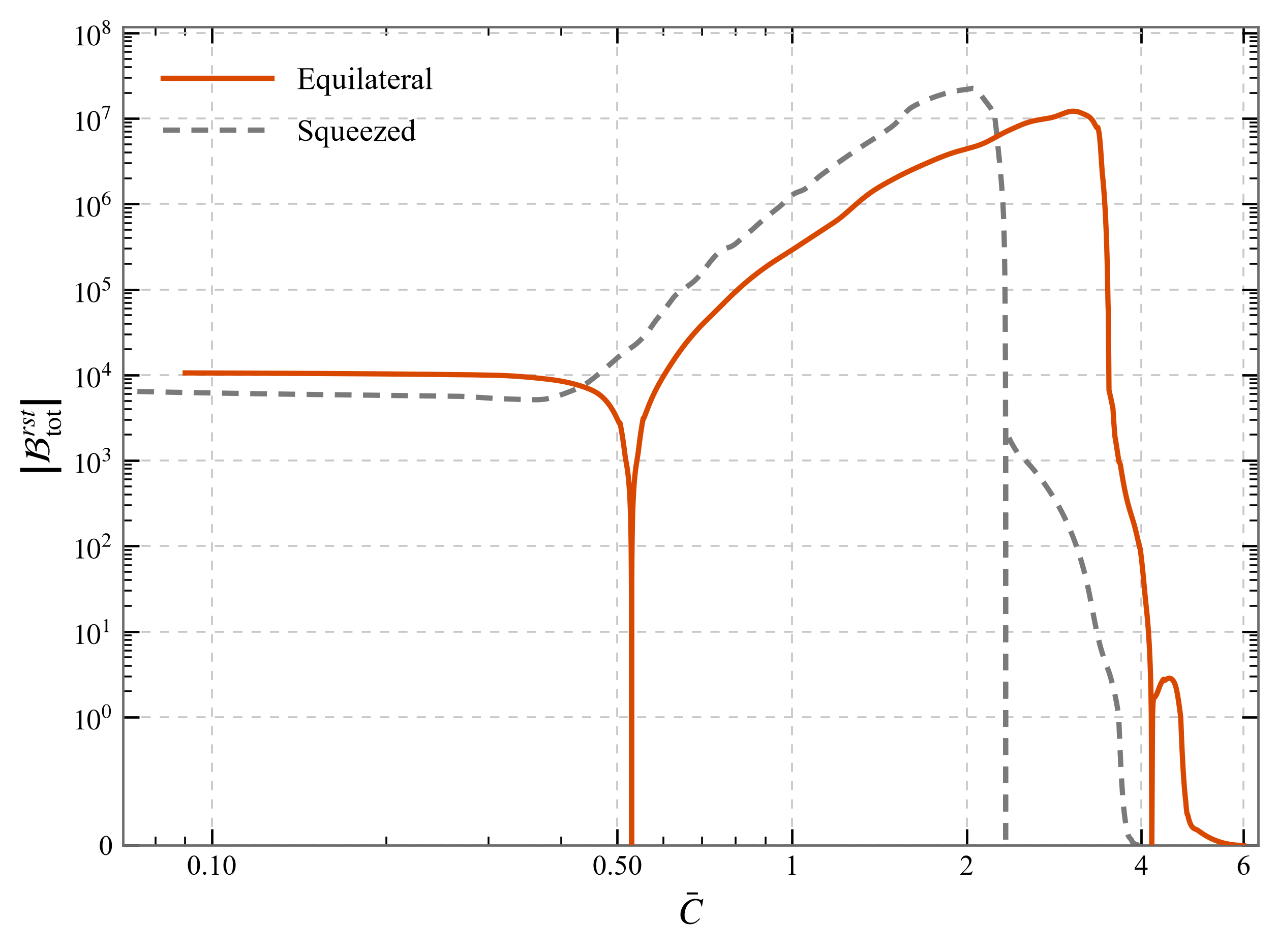}
	\caption{
		Absolute value of the total bispectrum scanned as a function of the perimeter $\bar C$. The red solid and gray dashed curves denote the equilateral and squeezed configurations, respectively.
	}
	\label{fig:equilateral_vs_squeezed}
\end{figure}

Another configuration corresponds to the squeezed limit, in which the three external momenta $k_i$ satisfy
$
	k_1\simeq k_2\gg k_3.
$. We consider a representative isosceles squeezed configuration,
$
	K_1=K_2\equiv K_{\rm short}
$, and fix the long-wavelength leg to $K_3=0.005$.
Figure~\ref{fig:polarizations_both}(\subref{fig:squeezed-polarizations}) shows two of the three independent polarization channels $RRR$ and $RRL$ of its bispectrum. In contrast to the equilateral case, the two polarization amplitudes are nearly degenerate over the full range, indicating that this configuration only weakly distinguishes the two circular polarization combinations. Starting from plateaus in $K_{short}<0.2$, they rise to peaks near the cutoff and then decrease, reach a local extremum, and finally vanish at $K_{short}=2$.

To examine the scale dependence of the preferred bispectrum configuration, Fig.~\ref{fig:equilateral_vs_squeezed} shows a scan of the total bispectrum with the normalized perimeter $\bar C$. Among the two benchmark configurations considered here, the larger amplitude switches between the equilateral and
squeezed cases as the perimeter varies: the equilateral configuration dominates both in the long wavelength regime and near the cutoff, whereas the squeezed configuration dominates for $\bar C=0.5\sim 2$. This behavior can be partly anticipated from the enhancement of the bispectrum near the cutoff: in the short wavelength regime, the perimeters at which the squeezed and equilateral configurations become dominant correspond, respectively, to the variable side reaching the cutoff. The equilateral dominance in the long wavelength regime, however, is difficult to attribute to a single mechanism. 

It is worth noting that this scale dependence differs from the
bispectrum behaviors reported for other induced-GW scenarios~\cite{
	Bartolo2019,
	PengFangGuo2024,
	HuFangZengGuo2025},
and therefore offers a possible discriminator among different GW
generation mechanisms.

Beyond presenting the full bispectrum in the PBH evaporation scenario, we use the $RRR$ channel to assess how effectively the bispectrum distinguishes a discrete source carrying the full hierarchy of Poisson statistics from a Gaussianized white-noise source containing only two-point information. The analytic results above show that their tensor bispectra differ by the $BB$ topology generated by the connected source three-point cumulant. To quantify its net effect on the absolute bispectrum amplitude, we define the symmetrically normalized amplitude difference
\begin{equation}
	\mathcal D_{\rm amp}
	\equiv
	\frac{
		\left|
		|\mathcal{B}_{\rm tot}^{rst}|
		-
		|\mathcal B_{PPP}^{rst}|\right|
	}{
		|\mathcal B_{\rm tot}^{rst}|
		+
		|\mathcal B_{PPP}^{rst}|
	}.
	\label{eq:D-amp}
\end{equation}

This quantity satisfies
$0\leq\mathcal D_{\rm amp}\leq1$. The scan results show that, after excluding enhancements induced by the near‑zero region of the bispectrum, the maximum value of the local relative correction is only
\begin{equation}
	\mathcal D_{\rm amp}
	\sim
	4\times 10^{-3},\quad \text{or}\quad\frac{|\mathcal B_{BB}^{rst}|}{|\mathcal B_{PPP}^{rst}|}\sim 7\times 10^{-3}.
\end{equation}
Further detailed examinations of the other curves show that the $BB$ contribution indeed has no appreciable impact on the locations and number of zeros or the configuration features of the spectrum.

This suggests that although the \(BB\) topology contains one additional
independent momentum integration, it does not provide a compensating
enhancement. Once the phase-space measure and the Poisson amplitude are combined consistently, cancellations in the
convolution kernel leave the $BB$ contribution subleading.

Thus, although the tensor bispectrum can formally distinguish an ideal Poisson source and a Gaussian‑sourced source sharing the same two‑point white‑noise spectrum, the discrimination efficiency is extremely low under our setup. The bispectrum exhibits weak practical sensitivity to the information encoded in the source’s intrinsic non‑Gaussianity.

\section{Conclusion and Discussion}
\label{sec:conclusion}

We have studied how the intrinsic non-Gaussianity of a Poisson type isocurvature source enters the power spectrum and bispectrum of induced GWs in the sudden PBH evaporation transition, with particular emphasis on evaluating the bispectrum. The relative ordering of the equilateral and benchmark squeezed
bispectrum amplitudes exhibits an equilateral--squeezed--equilateral
sequence as the scale changes. This differs from the poltergeist model, which is also enhanced by the abrupt transition eMD $\to$ lRD, yet adopts a nearly scale‑invariant scalar spectrum. It also differs from the scalar‑induced GWs in the pure RD era dominated by the equilateral configuration \cite{Bartolo2019}.
As another reference case, the primordial tensor bispectrum generated in standard single‑field inflation gets enhanced in the squeezed limit \cite{GaoEtAl2013}.
 This behavior therefore may help distinguish the discrete PBH scenario from other GW generation mechanisms. 

At the analytic level, the tensor power spectrum cannot distinguish a Gaussian white-noise source from a discrete Poisson source, whereas the bispectrum lifts this degeneracy. In the numerical results, however, the effect of the $BB$ contribution on the bispectrum remains subleading. This result reveals the distinction between the two types of non‑Gaussianity.
For the nonlinearity intrinsic to the GW generation process, even when the scalar source is Gaussianized, the quadratic induction relation still yields a substantial tensor bispectrum via Wick contractions.
In contrast, the channel transmitting the intrinsic non-Gaussianity of
the source is strongly suppressed by cancellations in the corresponding
convolution and remains subleading after the full statistical and
phase-space normalization is included. In other words, under our mechanism, the bispectrum mainly reflects the non‑Gaussianity originating from the former rather than the latter.

Our analysis extends beyond unclustered monochromatic PBHs to other rare discrete sources with approximately Poisson statistics, such as oscillons and Q-balls. If these objects decay rapidly after dominating the Universe, their number density fluctuations likewise generate a long-wavelength shot noise spectrum,
\(\mathcal P_S\propto K^3\), and can be treated approximately within the present sudden transition framework
\cite{LozanovSasakiTakhistov2025,SuiLiuCai2024}.
Extended PBH mass functions require generalizing the single amplitude
\(A_P\) and evaporation hypersurface used here~\cite{
	Papanikolaou2022,
	GouttenoireLeisterSchwaller2026},
whereas clustering, exclusion, or interactions introduce additional scale and exchanged-momentum dependence and may invalidate the selection rule
\cite{HeEtAl2024}.

A related limitation concerns the linear evolution of the initial
Poisson isocurvature perturbations.
During the PBH-dominated era, the PBH density contrast can grow to order unity or larger before evaporation, implying that sufficiently small-scale PBH fluctuations may enter the nonlinear regime~\cite{Papanikolaou2022,DomenechTrankle2025}. Following the linear treatment adopted in previous analyses, we nevertheless extrapolate the Poisson scalar statistics up to the discreteness cutoff $k_{\rm UV}$~\cite{PapanikolaouVenninLanglois2021,DomenechLinSasaki2021}.
Since
the corresponding nonlinear scale \(k_{\rm NL}\) may lie
below the discreteness cutoff \(k_{\rm UV}\), small scale scalar fluctuations may receive corrections from nonlinear PBH evolution and clustering. Related limitations have also been discussed in studies of the power spectrum~\cite{Papanikolaou2022,DomenechTrankle2025}, while consistently incorporating such nonlinear corrections into the PBH-discreteness bispectrum considered here lies beyond the scope of the present work. Our results should therefore be interpreted as predictions within the linear Poisson-isocurvature model. A more complete treatment of nonlinear PBH evolution may modify the ultraviolet scalar statistics and introduce additional uncertainty in the quantitative bispectrum amplitude and its detailed UV behavior.

\begin{acknowledgments}
This work is supported in part by the National Natural Science Foundation
of China Grants No.~12475067 and No.~12235019.
We also acknowledge the use of the HPC Cluster of ITP-CAS.
\end{acknowledgments}

\appendix

\section{Details of the Time-Kernel Integral}
\label{app:lRD-kernel}
This appendix gives the analytic method and result for the three-sine integral in the lRD time kernel. To discuss the resonant limit correctly, we first retain the finite upper limit $x_1$ in Eq.~\ref{eq:J-kernel-def} and take the late-time limit only after obtaining the analytic expression:
\begin{align}
 \mathcal J(a,b;x_1,x_3)
 ={}&\int_0^{x_1}
 \frac{dx_2}{x_2+x_3}
 \sin(x_1-x_2)
 \nonumber\\
 &\times\sin(a x_2)\sin(b x_2),
 \label{eq:app_three_sine_integral}
\end{align}
where
\begin{equation}
	a=c_su,\qquad b=c_sv,\qquad x_3=x_{\rm eva}/2.
	\label{eq:app_ab_def}
\end{equation}
To obtain the analytic result, we repeatedly use product-to-sum identities, including
\begin{equation}
	\sin(a x)\sin(b x)
	=
	\frac{1}{2}
	\left[
	\cos\bigl((a-b)x\bigr)
	-
	\cos\bigl((a+b)x\bigr)
	\right].
	\label{eq:app_product_to_sum_1}
\end{equation}
The product of three sine functions can then be reduced to sine and cosine functions with the four frequencies $1\pm a\pm b$. With $y=x_2+x_3$, the integral becomes a linear combination of
\(\int dy\,\sin(\lambda y+\varphi)/y\)
and
\(\int dy\,\cos(\lambda y+\varphi)/y\).
For real $\lambda\neq0$ and $y>0$, the corresponding real antiderivatives are
\begin{equation}
	\int \frac{\cos(\lambda y)}{y}\,dy
	=
	\operatorname{Ci}(|\lambda|y),
	\qquad
	\int \frac{\sin(\lambda y)}{y}\,dy
	=
	\operatorname{Si}(\lambda y),
	\label{eq:app_Ci_Si_standard}
\end{equation}
where the sign of the frequency is retained in the argument of $\operatorname{Si}$. For any fixed nonzero frequency $\lambda$, the late-time condition $|\lambda|x_1\gg1$ ensures that the upper endpoint lies in the large-argument regime. Since $\operatorname{Ci}(|\lambda|y)\to0$ and $\operatorname{Si}(\lambda y)\to(\pi/2)\operatorname{sgn}(\lambda)$ as $y\to\infty$, the integration upper limit can then be safely extended to infinity. Away from the resonance band, this condition is satisfied by all four frequency combinations at sufficiently late times.

Using the analytic result given in Eq.~(C4) of Appendix C of~\cite{DomenechLinSasaki2021}, an antiderivative with respect to $y$ can be written as
\begin{align}
	\mathcal F(y)={}&
	\frac14
	\Bigl\{
	\operatorname{Ci}\!\left(|1+a-b|y\right)
	\nonumber\\
	&\quad\times
	\sin\!\left[x_1+(1+a-b)x_3\right]
	\nonumber\\
	&+
	\operatorname{Ci}\!\left(|1-a+b|y\right)
	\nonumber\\
	&\quad\times
	\sin\!\left[x_1+(1-a+b)x_3\right]
	\nonumber\\
	&-
	\operatorname{Ci}\!\left(|1-a-b|y\right)
	\nonumber\\
	&\quad\times
	\sin\!\left[x_1-(-1+a+b)x_3\right]
	\nonumber\\
	&-
	\operatorname{Ci}\!\left(|1+a+b|y\right)
	\nonumber\\
	&\quad\times
	\sin\!\left[x_1+(1+a+b)x_3\right]
	\nonumber\\
	&+
	\cos\!\left[x_1+(1-a+b)x_3\right]
	\nonumber\\
	&\quad\times
	\operatorname{Si}\!\left[(-1+a-b)y\right]
	\nonumber\\
	&-
	\cos\!\left[x_1+(1+a-b)x_3\right]
	\nonumber\\
	&\quad\times
	\operatorname{Si}\!\left[(1+a-b)y\right]
	\nonumber\\
	&-
	\cos\!\left[x_1-(-1+a+b)x_3\right]
	\nonumber\\
	&\quad\times
	\operatorname{Si}\!\left[(-1+a+b)y\right]
	\nonumber\\
	&+
	\cos\!\left[x_1+(1+a+b)x_3\right]
	\nonumber\\
	&\quad\times
	\operatorname{Si}\!\left[(1+a+b)y\right]
	\Bigr\}.
	\label{eq:app_C4_result}
\end{align}
Under this late-time condition, Eq.~\eqref{eq:app_three_sine_integral} becomes
\begin{align}
	\mathcal J(a,b;x_1,x_3)
	\simeq{}&
	\int_0^{\infty}
	\frac{dx_2}{x_2+x_3}
	\sin(x_1-x_2)
	\nonumber\\
	&\times\sin(a x_2)\sin(b x_2)
	\nonumber\\
	={}&
	\mathcal F(\infty)-\mathcal F(x_3).
	\label{eq:app_definite_from_F}
\end{align}

A separate treatment is required in the resonance band, where the late-time condition above is not uniform. Within the physical integration region relevant here, the frequency combination that can approach zero is $1-a-b$. We define
\begin{align}
	\delta&\equiv1-a-b,
	\nonumber\\
	\delta=0
	&\quad\Longleftrightarrow\quad
	a+b=1,
	\nonumber\\
	&\hspace{4.2em}u+v=c_s^{-1}.
	\label{eq:app_resonance_condition}
\end{align}
This is the acoustic resonance of scalar-induced GWs. Extracting the terms proportional to the frequency $\delta$ from Eq.~\eqref{eq:app_C4_result} gives the finite-time resonant contribution
\begin{align}
	\mathcal J_{\rm res}
	={}&
	\frac14\Bigg\{
	-\sin(x_1+\delta x_3)
	\left[
	\operatorname{Ci}\!\left(|\delta|(x_1+x_3)\right)
	-
	\operatorname{Ci}\!\left(|\delta|x_3\right)
	\right]
	\nonumber\\
	&+
	\cos(x_1+\delta x_3)
	\left[
	\operatorname{Si}\!\left(\delta(x_1+x_3)\right)
	-
	\operatorname{Si}\!\left(\delta x_3\right)
	\right]
	\Bigg\}.
	\label{eq:app_resonant_contribution}
\end{align}
The small-argument expansions
\begin{align}
	\operatorname{Ci}(z)
	&=
	\gamma_E+\ln z+\mathcal O(z^2),
	\nonumber\\
	\operatorname{Si}(z)
	&=
	z+\mathcal O(z^3)
	\label{eq:app_Ci_Si_small_argument}
\end{align}
show that, as $\delta\to0$ at fixed $x_1$,
\begin{align}
	&\operatorname{Ci}\!\left(|\delta|(x_1+x_3)\right)
	\nonumber\\
	&\quad
	-
	\operatorname{Ci}\!\left(|\delta|x_3\right)
	\longrightarrow
	\ln\!\left(\frac{x_1+x_3}{x_3}\right),
	\nonumber\\
	&\operatorname{Si}\!\left(\delta(x_1+x_3)\right)
	\nonumber\\
	&\quad
	-
	\operatorname{Si}\!\left(\delta x_3\right)
	\longrightarrow0.
	\label{eq:app_resonant_differences}
\end{align}
Consequently, the exact resonance remains finite at finite observation time:
\begin{equation}
	\mathcal J_{\rm res}(\delta=0)
	=
	-\frac14\sin x_1
	\ln\!\left(\frac{x_1+x_3}{x_3}\right).
	\label{eq:app_exact_resonance}
\end{equation}

The late-time limit is non-uniform near resonance. For fixed nonzero $\delta$, the upper-endpoint $\operatorname{Ci}$ and $\operatorname{Si}$ functions can be replaced by their large-argument asymptotic limits only when
\[
	{
	|\delta|(x_1+x_3)\gg1.
	}
\]
Even for large $x_1$, this condition fails sufficiently close to $\delta=0$. Thus, the limits $\delta\to0$ and $x_1\to\infty$ do not commute. Taking the late-time limit first at fixed $\delta\neq0$ gives the familiar logarithmic behavior $\operatorname{Ci}(|\delta|x_3)\sim\ln|\delta|$, whereas taking $\delta\to0$ first at finite $x_1$ gives the finite logarithmic ratio in Eq.~\eqref{eq:app_exact_resonance}. Close to resonance, the finite-time expression shows explicitly how the apparent logarithmic singularity is regulated by the finite upper limit.

\section{\texorpdfstring{Connected-Block Decomposition and TT Selection Rules}{Connected-Block Decomposition and TT Selection Rules}}
\label{app:TT-selection}
\begingroup

We give the momentum-space steps behind the selection rule used in Sec.~\ref{sec:formalism}. For each tensor vertex, let $S_{ip}\equiv S(\mathbf p_i)$ and $S_{iq}\equiv S(\mathbf q_i)$, where $\mathbf q_i=\mathbf k_i-\mathbf p_i$. The initial source has zero mean, so the connected partitions of its four-point moment have block sizes $2+2$ and $4$, denoted by $PP$ and $T$. Those of its six-point moment have sizes $2+2+2$, $3+3$, $2+4$, and $6$, denoted by $PPP$, $BB$, $PT$, and $C_6$. Before the tensor projection, these classes contain respectively $3$ and $1$ partitions at four-point order, and $15$, $10$, $15$, and $1$ partitions at six-point order.

The reduced source power spectrum $P_S$ and bispectrum $B_S$ were defined in Eq.~\eqref{eq:B_connected_def}. To specify the remaining symbols without introducing a separate general cumulant notation, define the connected trispectrum $T_S$ and six-point spectrum $C_{6,S}$ by
\begin{align}
 \left\langle\prod_{a=1}^{4}S(\boldsymbol\ell_a)\right\rangle_c
 ={}&(2\pi)^3\delta_{\rm D}^{(3)}
 \!\left(\sum_{a=1}^{4}\boldsymbol\ell_a\right)
 \nonumber\\
 &\times T_S(\boldsymbol\ell_1,\ldots,\boldsymbol\ell_4),
 \label{eq:app-TS-def}\\
 \left\langle\prod_{a=1}^{6}S(\boldsymbol\ell_a)\right\rangle_c
 ={}&(2\pi)^3\delta_{\rm D}^{(3)}
 \!\left(\sum_{a=1}^{6}\boldsymbol\ell_a\right)
 \nonumber\\
 &\times C_{6,S}(\boldsymbol\ell_1,\ldots,\boldsymbol\ell_6).
 \label{eq:app-C6S-def}
\end{align}
For the homogeneous Poisson source of Eq.~\eqref{eq:poisson-Npoint}, their reduced spectra are
\begin{equation}
 T_S=A_P^3W_4,
 \qquad C_{6,S}=A_P^5W_6,
 \label{eq:app-T-C6-poisson}
\end{equation}
where each window $W_n$ is the product of one cutoff factor per scalar momentum. In particular, apart from the momentum-conserving delta function, these connected blocks contain no mixed angular dependence between distinct scalar momenta. This factorization property is essential to the following argument.

The basic TT identity concerns a tensor vertex whose internal momentum is unconstrained by the other vertices. Its remaining integrand depends only on $p=|\mathbf p|$ and $\mu=\hat{\mathbf k}\cdot\hat{\mathbf p}$, and therefore has the form
\begin{align}
 \mathscr L_\lambda(\mathbf k)
 &\equiv \int\dd^3p\,
 \Pi_\lambda(\mathbf k,\mathbf p)F(p,\mu)
 \nonumber\\
 &=\frac{e^\lambda_{ij}(\hat{\mathbf k})}{k^2}
 \int\dd^3p\,p^ip^jF(p,\mu).
 \label{eq:app-single-leg}
\end{align}
Rotational symmetry about $\mathbf k$ implies
\begin{equation}
 \int\dd^3p\,p^ip^jF(p,\mu)
 =A\delta^{ij}+B\hat k^i\hat k^j.
 \label{eq:app-axisymmetric-tensor}
\end{equation}
Since $e^{\lambda}_{ii}=0$ and $e^\lambda_{ij}\hat k^j=0$, Eq.~\eqref{eq:app-single-leg} gives $\mathscr L_\lambda(\mathbf k)=0$. Here $F$ may include the time kernel and the cutoff factors belonging to this tensor vertex.

First consider the connected $T$ contribution to $\langle hh\rangle$. Its only momentum-conserving delta function reduces to
\begin{equation}
 \delta_{\rm D}^{(3)}
 (\mathbf p_1+\mathbf q_1+\mathbf p_2+\mathbf q_2)
 =\delta_{\rm D}^{(3)}(\mathbf k_1+\mathbf k_2).
 \label{eq:app-T-delta}
\end{equation}
After the overall external delta function is removed, $\mathbf p_1$ and $\mathbf p_2$ are independent. Equation~\eqref{eq:app-T-C6-poisson} then lets the two vertex integrals factorize, yielding
\begin{equation}
 \langle h^r_{\mathbf k_1}h^s_{\mathbf k_2}\rangle_T
 \propto
 \mathscr L_r(\mathbf k_1)\mathscr L_s(\mathbf k_2)=0.
 \label{eq:app-T-vanishes}
\end{equation}
For the $PP$ class, a pairing within one tensor vertex imposes $\delta_{\rm D}^{(3)}(\mathbf k_i)$ and does not contribute at nonzero external momentum. The other two pairings connect the two vertices and survive this selection rule.

For the $PT$ class in $\langle hhh\rangle$, a two-point block that pairs the two fields of one tensor vertex again imposes $\delta_{\rm D}^{(3)}(\mathbf k_i)$. If the pair instead joins distinct tensor vertices, the third tensor vertex contributes both of its source fields to the four-point block. For example, take $P(S_{1p},S_{2p})\,T(S_{1q},S_{2q},S_{3p},S_{3q})$. Its two delta functions obey
\begin{align}
 &\delta_{\rm D}^{(3)}(\mathbf p_1+\mathbf p_2)
 \delta_{\rm D}^{(3)}
 (\mathbf q_1+\mathbf q_2+\mathbf p_3+\mathbf q_3)
 \nonumber\\
 &\quad=
 \delta_{\rm D}^{(3)}(\mathbf p_1+\mathbf p_2)
 \delta_{\rm D}^{(3)}
 (\mathbf k_1+\mathbf k_2+\mathbf k_3).
 \label{eq:app-PT-deltas}
\end{align}
The momentum $\mathbf p_3$ is unconstrained, so its integral is $\mathscr L_t(\mathbf k_3)=0$. The same reasoning applies to every cross-vertex pairing; hence all $15$ $PT$ partitions vanish.

The connected $C_6$ block likewise supplies only the overall external delta function,
\begin{equation}
 \delta_{\rm D}^{(3)}
 \!\left(\sum_{i=1}^{3}(\mathbf p_i+\mathbf q_i)\right)
 =\delta_{\rm D}^{(3)}
 \!\left(\sum_{i=1}^{3}\mathbf k_i\right).
 \label{eq:app-C6-delta}
\end{equation}
The three internal momenta are independent after this delta function is removed. The Poisson window $W_6$ factors by tensor vertex, so the contribution is proportional to $\mathscr L_r(\mathbf k_1)\mathscr L_s(\mathbf k_2)\mathscr L_t(\mathbf k_3)=0$.

Of the $10$ unordered $BB$ partitions into two triples, $6$ vanish. In each of these, one block contains both source fields from a tensor vertex. For example, in $B(S_{1p},S_{1q},S_{2p})\,B(S_{2q},S_{3p},S_{3q})$, the first block constrains $\mathbf k_1+\mathbf p_2$ but not $\mathbf p_1$ itself. The second block also does not constrain $\mathbf p_1$. Its integral at the first tensor vertex therefore separates as $\mathscr L_r(\mathbf k_1)=0$. The same argument applies to the other five partitions after relabeling the vertices and source fields.

The remaining $2^3/2=4$ partitions place one source field from each tensor vertex in each three-point block. For example, $B(S_{1p},S_{2p},S_{3p})\,B(S_{1q},S_{2q},S_{3q})$ gives
\begin{align}
 &\delta_{\rm D}^{(3)}
 (\mathbf p_1+\mathbf p_2+\mathbf p_3)
 \delta_{\rm D}^{(3)}
 (\mathbf q_1+\mathbf q_2+\mathbf q_3)
 \nonumber\\
 &\quad=
 \delta_{\rm D}^{(3)}
 (\mathbf p_1+\mathbf p_2+\mathbf p_3)
 \delta_{\rm D}^{(3)}
 (\mathbf k_1+\mathbf k_2+\mathbf k_3).
 \label{eq:app-BB-deltas}
\end{align}
Thus, after the external delta function is factored out, one internal closure condition remains: $\mathbf p_3=-\mathbf p_1-\mathbf p_2$. The third polarization factor consequently depends on both remaining integration momenta, and no single-leg angular integral separates. The TT identity therefore does not force these four partitions to vanish.

Finally, a $PPP$ partition contributes at nonzero external momenta only if no pair lies within one tensor vertex. Every surviving pairing then links each vertex once to each of the other two, forming a cycle. There are $2^3=8$ such pairings among the $15$ pair partitions. Their pairwise momentum constraints leave one common internal momentum, as in Eq.~\eqref{eq:Bh-PPP-dimensionful}, so none of the tensor-vertex integrations has the separable form of Eq.~\eqref{eq:app-single-leg}. These results establish the $PP$ and $PPP+BB$ decompositions used in the main text for the ideal Poisson source.
\endgroup

\end{document}